\documentstyle[twoside,ijmpa,epsfig]{article}
\def\qed{\hbox{${\vcenter{\vbox{                        
   \hrule height 0.4pt\hbox{\vrule width 0.4pt height 6pt
   \kern5pt\vrule width 0.4pt}\hrule height 0.4pt}}}$}}

\def\bsc{{\sc a\kern-6.4pt\sc a\kern-6.4pt\sc a}}       
\def\bflatex{\bf L\kern-.30em\raise.3ex\hbox{\bsc}\kern-.14em
T\kern-.1667em\lower.7ex\hbox{E}\kern-.125em X}

\begin{document}
\runninghead{Search for DCC in relativistic heavy-ion collisions : Possibilities and Limitations} {B. Mohanty, T.K. Nayak, D.P. Mahapatra  and Y.P. Viyogi}
\normalsize\textlineskip
\thispagestyle{empty}
\setcounter{page}{1}

\copyrightheading{}                     

\vspace*{0.88truein}

\fpage{1}
\centerline{\bf 
Search for DCC in relativistic heavy-ion collisions : 
Possibilities and Limitations}
\vspace*{0.37truein}

\centerline{\footnotesize 
  B.~Mohanty$^{1}$\footnote{Corresponding author, e-mail: bmohanty@veccal.ernet.in
    }, T.K. Nayak$^{1}$, D.P. Mahapatra$^{2}$ and Y.P. Viyogi$^{1}$}
\vspace*{0.015truein}
\centerline{\footnotesize\it $^{1}$Variable Energy Cyclotron Centre, Kolkata, 700-064 India}
\baselineskip=10pt
\centerline{\footnotesize\it $^{2}$Institute of Physics, Bhubaneswar, 751-005 India}

\vspace*{0.21truein}

\abstracts{
   The experimental observation of disoriented chiral condensate 
   is affected due to various physical and detector related effects. 
   We study and quantify the strength of the experimental signal,
   ``neutral pion fraction'' within the framework of a simple DCC model,
   using the analysis methods based on the multi-resolution discrete 
   wavelet technique and by evaluating the signal to background ratio. 
   The scope and limitations of DCC search in heavy-ion collision 
   experiments using various combination of detector systems are investigated.\\}{}{}

\vspace*{1pt}\textlineskip      
\vspace*{-0.5pt}

\noindent {PACS-key : 25.75.Gz,13.40.-f and Heavy-ion collisions, Disoriented chiral condensates } \\

\section{Introduction}
\label{sec:1}

   Disoriented chiral condensates (DCC), localized in phase space,
   have been predicted to be formed in high energy heavy-ion 
   collisions when the chiral symmetry is restored at high 
   temperatures~\cite{raj1}. The formation of DCC leads to  
   large event-by-event fluctuation in the neutral pion fraction ($f$), 
   defined as 
\begin{equation}
f = N_{\pi^0}/N_{\pi}
\label{f_def}
\end{equation}
\noindent{where $N_{\pi^0}$ and $N_{\pi}$ are multiplicities of neutral 
and total pions respectively.} The probability distribution of $f$ inside 
a DCC domain is given by~\cite{blai1,bjor1} 
\begin{equation}
 P(f) =  \frac{1}{2\sqrt{f}}  
\label{f_prob}
\end{equation}
   For normal events the distribution of $f$ is a binomial 
   peaking at $\frac{1}{3}$. The formation of DCC domains gives rise to 
   pion multiplicity fluctuations resulting in the neutral pion fraction 
   deviating significantly from $\frac{1}{3}$. Experimentally $\pi^0$ is 
   detected by its decay product which are photons. Since most of the charged 
   particles detected in an event are charged pions and similarly majority
   of the photons originate from $\pi^0$ decay, DCC formation  would thus 
   lead to event-by-event correlated fluctuations in the 
   number of charged particles and photons in a given phase space.

   Many experiments have attempted to look for DCC and several
   others are coming up. Experiments so far carried out are based
   on both cosmic ray studies and accelerators for colliding hadrons and 
   heavy-ions at high energy. The JACEE collaboration had reported the 
   occurrence of a typical anti-CENTAURO event (high $f$ values)~\cite{jacee}.
   The mountain top experiments, carried out at Mt. Chacaltaya in Bolivia and
   Mt. Fuji in Japan, had reported CENTAURO type events 
   (low $f$ values)~\cite{jacee}.   
   Recently there have been encouraging results from other cosmic ray 
   experiments at Pamir~\cite{augusto}. In the accelerator based experiments, 
   UA1~\cite{ua1} and UA5~\cite{ua5}, have done extensive searches 
   for CENTAURO events at the CERN-SPS ($\bar{pp}$) 
   collider. However no indication of CENTAURO events was observed 
   up to $\sqrt{s} = 900$ GeV. D0~\cite{d0}, CDF~\cite{cdf} and 
   Minimax~\cite{minimax} experiments at Fermilab have reported 
   null results. The WA98~\cite{WA98-12} and NA49~\cite{NA49pt} 
   heavy-ion experiments at CERN-SPS have reported upper limits on DCC 
   production at SPS. While the analysis of the WA98 experiment 
   was based on a systematic study
   of photon and charged particle multiplicity correlation using the data 
   from a preshower photon multiplicity detector (PMD) and a silicon pad
   multiplicity detector (SPMD) for charged particles, the NA49 experiment
   tried to obtain the limits on DCC-type fluctuations based on information 
   from charged particles only.

   There now exist many powerful analysis techniques to look for DCC, 
   based essentially on the study of fluctuation  
   in the multiplicity of charged particles and photons
   \cite{minimax,WA98-12,dccflow,dccstr,dccprobe,dccphi}. 
   However, the absence of a reliable dynamical DCC model and the lack 
   of information on various factors affecting the observation of DCC
   (such as the probability of occurrence of DCC in a
   reaction at a particular energy, number of possible DCC domains 
   in an event, size of the
   domains, number of pions emitted from the domains and the
   interaction of the DCC pions with the rest of the system
   \cite{raj1,asakawa,horm1,randrup})  
   make experimental search of domains of DCC difficult. In a recent
   theoretical study~\cite{krzy}, within the limitations of
   a simple model of DCC, the probability of formation of DCC at
   SPS has been predicted to be of the order of $10^{-3}$ which is
   close to the upper limit set by an involved analysis of the WA98 data
   ~\cite{WA98-12}. 

   For all these experimental techniques to succeed in detecting 
   the DCC signal with very low probability of occurrence, it is necessary to 
   understand, from an experimental point of view, all the factors that affect
   the detection of DCC domain during its transition from the time of 
   formation to the time of detection. We study this transition by the help 
   of a simple DCC model. The analysis is carried out using the method based 
   on discrete wavelet transformation (DWT) and looking at the signal to 
   background ratio.

   Possible factors that affect the DCC signal are :
   the presence of multiple domains of DCC, $\pi^{0}$ decay, increasing
   particle multiplicity as we go from SPS to RHIC and LHC energies, and 
   the detector related effects, such as  efficiency and  purity of particle 
   detection and use (or lack) of $p_{T}$ information of the 
   detected particles. 
   All these effects are discussed in this paper. This is important from the 
   point of view that the current and future heavy-ion experiments at RHIC 
   and LHC have kept DCC search high on their agenda. At RHIC, STAR 
   experiment plans to look for DCC using the combination of photon 
   multiplicity detector (PMD) and the electromagnetic calorimeter (EMCAL) 
   for photon detection along with the time projection chamber (TPC) and 
   forward time projection chamber (FTPC) for charged particle detection
   ~\cite{starnim}. The PHENIX experiment has also demonstrated their 
   capability to study isospin fluctuations, essentially through charged 
   particle measurements~\cite{phenixqm}.
   At LHC, the ALICE experiment 
   will have the combination of a PMD and a forward charged 
   particle multiplicity detector (FMD) as well as an 
   electromagnetic calorimeter (PHOS) with the time projection chamber (TPC)
   in a different rapidity region for such a study~\cite{alicetp}.
   In the present article we concentrate on the possibilities 
   and limitations of
   searching for DCC in experiments  using the measurement of 
   charge-to-neutral ratio where the neutral particle
   (photons primarily from $\pi^{0}$ decay) is handled using a preshower 
   detector. We will also briefly discuss other detector combinations.

   The paper is organized as follows. In the next section we discuss a simple
   model of DCC and the methods of analysis used for the present study. In the
   subsequent sections we discuss the various factors that affect the DCC
   search using the discrete wavelet transformation technique (DWT).
   In Section 5 we present the results on the signal to 
   background ratio. 
   Finally we summarize with a discussion on the various 
   limitations and possibilities for DCC search in heavy-ion collisions.

\section{DCC model and method of analysis}
\label{sec:2}
\subsection{DCC model and simulated events}
\label{subsec2:1}

   In the absence of a reliable dynamical model 
   where the effect of the formation of DCC could be simulated at 
   an early stage of the reaction, we take a simple DCC model where
   the generic particle production is governed by the VENUS event 
   generator~\cite{venus} and the DCC pions are introduced at 
   the freeze-out stage. The goal is to observe the effect of DCC 
   domains on the measured quantities, assuming DCC pions survive 
   till the freeze-out time. This assumption is justified by the recent 
   calculations~\cite{koch}.  

   The following are the main features of the model.

\begin{enumerate}

\item DCC-type fluctuations : For DCC simulation in a given domain, 
      the identity of charged pions taken pairwise is changed to 
      neutral ones  and vice-versa~\cite{WA98-12,dccstr},  
      following the DCC probability as given by Eqn.~(\ref{f_prob}).

\item Domain size :  The size of a domain is defined 
   in terms of its extent in pseudo-rapidity ($\eta$) and azimuthal angle 
   ($\phi$).  We have chosen a  DCC domain with $\Delta \eta$ = 1 
   and then varied the extent in $\phi$ (denoted as $\Delta\phi$)
   to see the effect on observed quantities.  

\item Number and $p_{T}$ spectrum of DCC pions :
   For simplicity all the pions inside a chosen domain are considered to 
   be DCC-type.  
   There are theoretical calculations which suggest that DCC pions may 
   have low $p_{T}$ and DCC formation may lead to enhancement in the 
   production of low $p_{T}$ pions~\cite{Kaga}. In order to study the 
   usefulness
   of $p_{T}$ information of particles, we consider two cases -
   (a)  only pions with $p_{T}$ ($\le 150$ MeV) are DCC pions in a given 
    domain, and (b) a number of pions (depending on the domain volume) 
    of $p_{T}$ $\le 150$ MeV having DCC origin are added to
   the generic sample of pions in a given domain. 

\item $\pi^{0}$ decay and photon fraction :
  After introducing DCC-type fluctuations the $\pi^{0}$'s are allowed
 to decay. The neutral pion fraction then gets modified to $photon$ 
 $fraction$ given as,
\begin{equation}
f^{\prime} = (N_{\mathrm \gamma}/2)/(N_{\mathrm \gamma}/2 + N_{\pi^{\pm}})
\label{new_f}
\end{equation}
   where  $N_{\mathrm \gamma}$ is the photon multiplicity and 
  $N_{\pi^{\pm}}$ is the total multiplicity of charged pions. 

\item Percentage of events being DCC-type :
The sets of events with DCC will be referred to as ``nDCC'' events.
The ensemble of nDCC events may or may not have all the events of DCC type.
nDCC event ensembles with varying percentage of DCC-type events were
generated to study the sensitivity of various physical and detector related
effect to the observation of DCC.

\end{enumerate}

\subsection{Charged particle and photon multiplicity detectors}
\label{subsec2:2}

   We assume a hypothetical photon detector placed in the forward region
   and covering full azimuth in one unit of pseudo-rapidity. 
   The detected photon sample has some admixture of charged particles due
   to the limitations of photon-hadron discrimination algorithm. The
   relation between the number of photons incident on the detector
   ($N_\gamma$) and the number of identified photon-like 
   candidates ($N_{\gamma-like}$)
   is given by ~\cite{WA98-9}
\begin{equation}
N_{\mathrm \gamma-like}  = \epsilon N_{\mathrm \gamma}/p
\label{pur_eqn}
\end{equation}
\noindent {where  $\epsilon$ is the photon 
counting efficiency and $p$ is the purity of the detected photon sample.

For the present study the photon detector is assumed to have 
realistic photon counting 
efficiency of $70\% \pm 5\%$ and purity of photon sample of 
$70\% \pm 5\%$ ~\cite{wa98nim,starpmd,alicepmd,subnim}. The efficiency
and purity values are varied to study the role of detector limitation in the
observation of DCC.

   The charged particles are detected using a hypothetical 
   charged particle detector with an 
   efficiency of $90\% \pm 5\%$~\cite{Lin,ftpc}. It is assumed that there is
   no contamination in the detected charged particle sample (the effect of
   $\delta$-rays are ignored). The charged particle detector is assumed to
   have complete overlap with the photon detector in $\eta-\phi$ coverage.

   For this work it is assumed that the DCC domain has a complete overlap 
   with the detectors. It may be mentioned that it is not easy to 
   experimentally measure the DCC domain size, if it has been formed in the 
   reaction. It is also difficult to find out experimentally whether the DCC
   domain formed has a full or partial overlap with the detectors.

\subsection{Discrete Wavelet method}
\label{subsec2:3}

   Multi-resolution discrete wavelet technique (DWT) has been 
   used to look for bin-to-bin fluctuations in charged particle 
   and photon multiplicity distributions of the simulated 
   events~\cite{WA98-12,dccstr,huang}. 
   The DWT technique has the beauty of 
   analyzing a distribution of particles at different length scales with 
   the ability of finally picking up the right scale at which there is a 
   fluctuation. This method has been utilized very successfully in many
   fields including image processing, data compression, turbulence, 
   human vision, radar and earthquake prediction. 
   This technique has been suitably adopted here to search for 
   bin to bin fluctuation in the charged particle and photon multiplicity 
   distributions.

   The analysis has 
   been carried out by making $2^{j}$ bins in $\phi$  where $j$ is the 
   resolution scale. The input to the DWT analysis is a sample 
   distribution function given by Eqn.~\ref{new_f} at the highest 
   resolution scale, $j_{max}$. The analysis has been carried out using the 
   D-4 wavelet basis~\cite{d4}. It may be mentioned that there are several
   families of wavelet bases distinguished by the number of coefficients and
   the level of iteration; we have used the frequently employed D-4 wavelet
   basis~\cite{d4,amara}.

\begin{figure}
\begin{center}
\includegraphics[scale=0.65]{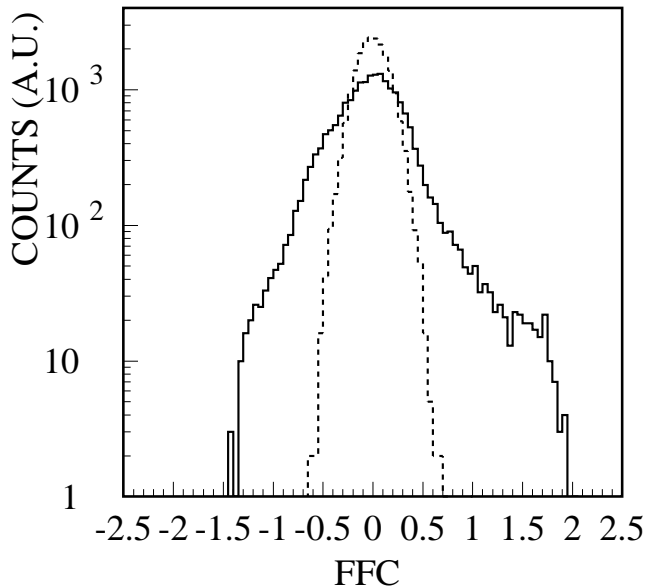}
\vskip -5.0cm
\caption {\label{ffcdist}
 FFC distribution for simulated generic VENUS (dotted line) and pure
 DCC-like events (solid line). The rms deviation of FFC distribution for
 DCC-like events is wider than those for VENUS events.
}
\end{center}
\end{figure}

   The output of the DWT analysis consists of a set of wavelet or father 
   function coefficients (FFC's) at each scale, $j = 1, ...(j_{max} - 1)$.
   It is these FFC's at a given scale that carry information of 
   fluctuation at higher scales. All the analysis have been carried out 
   by taking $j_{max} = 5$. 
   Due to the completeness and orthogonality of the DWT basis, there is
   no information loss at any scale.

   The FFC distribution of normal events is a 
   Gaussian while the distribution for events  having DCC-type 
   fluctuations has non-Gaussian shapes~\cite{WA98-12,dccflow,dccstr}.
   In Fig.~\ref{ffcdist} we show the typical FFC distribution for pure 
   generic and pure DCC-like events at scale $j = 1$. The DCC-like events have 
   a DCC domain of $\Delta \phi$ = $90^{\circ}$ in all events. 
   We find that the FFC distribution for DCC-like events is broader 
   in comparison to that for generic events. The amount of broadening depends 
   on various features associated with DCC domains,
   In fact, one can show that 
   there is pile up of a large number of events within the width of the 
   generic distribution with decrease in the fraction of DCC-like events. 

   For comparison of these distributions we use the root mean square (rms)
   deviations instead of the usual standard deviation.
   A direct comparison of the rms
   deviation of FFC distributions for different cases will yield 
   information about the fluctuations. To quantify this effect 
   properly, we define a quantity called strength of DCC signal
   ($\zeta$)  as,
\begin{equation}
 \zeta = \frac{\sqrt{(s_{\rm X}^2 - s_{\rm N}^2)}}{s_{\rm N}}
\label{str_para}
\end{equation}
  where $s_{\rm N}$ is the rms deviation of the FFC distribution 
  for normal event set and $s_{\rm X}$ is  the rms deviation 
  for the different nDCC event sets.
  The quantity $\zeta$ also reflects the detectability of the DCC signal.
  The statistical error on $\zeta$ has been calculated by varying the 
  number of events and it was found to be $\sim 0.1$ for the case when all 
  pions in a domain are DCC pions.

\begin{figure}
\begin{center}
\includegraphics[scale=0.4]{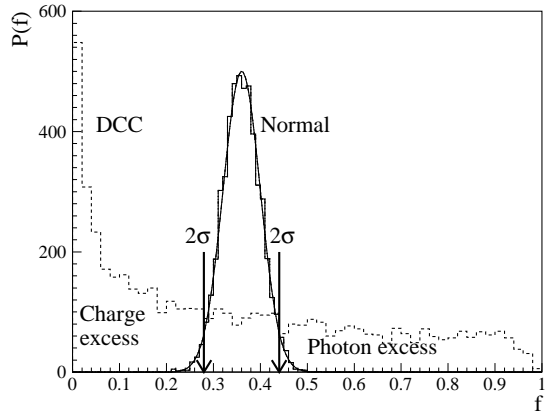}
\caption {\label{fndist}
Schematic diagram of the $f$ distribution of DCC events 
(dashed lines) and the normal Gaussian distribution of generic events 
(solid line) peaking at 1/3. Also shown are the regions of photon excess and
charge excess and the region within to 2$\sigma$ of the normal 
distribution.
}
\end{center}
\end{figure}

\subsection{Signal and Background of DCC events}
\label{subsec2:4}

  In a generic particle production mechanism, because of isospin
  conservation in strong interaction, the production of $\pi^{0}$,
  $\pi^{+}$ and $\pi^{-}$, are equally probable. This mechanism
  leads to a binomial distribution of neutral pion fraction ($f$)
  peaking at $\frac{1}{3}$. For a large number of events having high 
  multiplicity of pions produced, as in heavy-ion collisions, this 
  distribution for normal events can be approximated as a 
  Gaussian with mean at $\frac{1}{3}$. Fig.~(\ref{fndist}) displays the $f$ 
   distribution of pure DCC events and pure generic events.
  Since the DCC events lying in the overlap region of the two distributions 
  are difficult to separate, it may be advisable in some analysis
  to exclude a region around the  generic peak. We define :

\begin{itemize}

\item   Signal (S)      =  Number of events beyond $ f = f_{N}^{peak} 
                                                     ~\pm~2\sigma_{N}$
                           in $f$ distribution of DCC sample,

\item    Background (B) =  The number of events beyond
                           $ f = f_{N}^{peak}~\pm~2\sigma_{N}$ 
                           in the $f$ distribution of normal sample,

\end{itemize}
where, $f_{N}^{peak}$ is the peak value and $\sigma_{N}$ is the standard 
deviation of the $f$ distribution of normal events respectively.

Here we separately consider the signal in two regions :
 $charge-excess$ events in the left part  and 
$photon-excess$ events in the right part of the Fig.~(\ref{fndist}).

With the analysis methods mentioned we now study the various factors
that affect the DCC signal.

\section{Physical effects}
\label{sec:3}

\begin{figure}
\begin{center}
\includegraphics[scale=0.4]{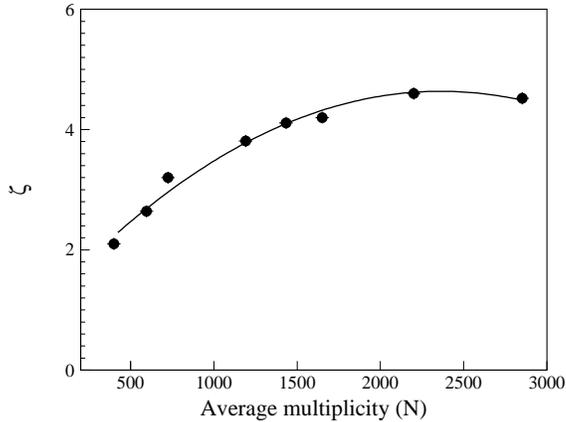}
\caption {\label{multi}
Variation in the strength of DCC signal
with multiplicity in a domain of $\Delta \phi$ = $90^{\circ}$ and 
$\Delta \eta$ = 1. $N$
is the average of the sum of mean photon and mean charged particles 
in a set of events. All the events are DCC events. The solid line is
proportional to $\sqrt N$.
}
\end{center}
\end{figure}

\subsection{Effect of Multiplicity}
\label{sec:3.1}

The effect of higher multiplicity is to reduce the event-by-event
statistical fluctuation associated with the multiplicity of photons 
and charged particles. A simple estimate of strength of DCC-type
fluctuation shows that for a DCC domain of $\Delta \phi$ = $90^{\circ}$  
and  $\Delta \eta$ of one unit, the strength of DCC signal increases 
with increase in average multiplicity of photons and charged particles. 
Fig.~(\ref{multi}) shows the variation 
of strength ($\zeta$) of DCC signal with multiplicity, where $N$ is the
average of the mean number of photon and mean number of 
charged particles in a set
of events for one unit in $\eta$. One observes that the increase is 
almost like a  $\sqrt{N}$ effect. This is probably a consequence of going
from a narrow Gaussian distribution to a wider $1/2\sqrt{f}$ distribution.
It may be mentioned that typical values of $N$ in central 
collisions for $\Delta \eta$ $\sim$ 1, 
for the WA98 experiment at SPS (combination of PMD and SPMD
detector for charged particle detection) was 320, for STAR (PMD + FTPC) 
will be about 500 and ALICE (PMD + FMD) will be about 2800 (for an assumed 
pseudo-rapidity density of $\sim$ 8000 at mid rapidity). 

\subsection{Effect of $\pi^{0}$ decay}
\label{sec:3.2}

\begin{figure}
\begin{center}
\includegraphics[scale=0.4]{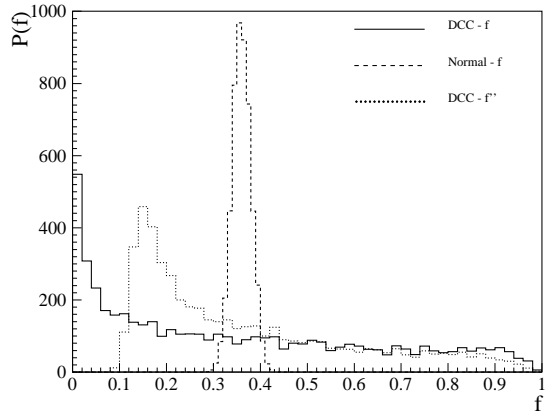}
\caption {\label{domain_deccay}
Typical $f$ distributions inside a DCC domain before (solid line) 
and after decay (dotted line) of $\pi^{0}$ ($f^{\prime \prime}$). 
Also shown for comparison is the $f$ distribution for normal
events (dashed line).
}
\end{center}
\end{figure}

  The formation of DCC leads to large event-by-event fluctuation
  in neutral pion fraction ($f$).  To observe this in an ideal situation 
  one has to count the $\pi^{0}$'s 
  event-by-event. But the $\pi^{0}$'s decay into photons 
  by the time they reach the detector. 

  As a result of decay, photons coming from DCC $\pi^{0}$'s
  inside a given DCC 
  domain may move out of the domain (depending on the momentum) 
  before they are detected. Also photons from outside the DCC domain
  will enter the $\eta-\phi$ phase space of the domain. 
  Both these effects will result in dilution of the strength of the signal.
  Considering the effect of $\pi^{0}$ decay, we modify Eqn.~(\ref{new_f}) as -
  
\begin{equation}
f^{\prime \prime} = (N_{\mathrm \gamma}/2 \pm \delta_{\gamma})/ (N_{\mathrm \gamma}/2 \pm  \delta_{\gamma}+ N_{\mathrm ch}),
\label{decay_eqn}
\end{equation}
  where $\delta_{\gamma}$ is the resultant number of photons that get removed
  from or added into the  DCC domain and $N_{\mathrm ch}$ is the total number
  of charged pions.  For multiplicity 
  detectors (e.g SPMD in WA98, FMD in ALICE) which do not have the 
  capability of particle identification, $N_{\mathrm ch}$ denotes 
  the multiplicity of all charged particles.

  Since $\delta_{\gamma}$ is a non zero number, the 
  possibility of observing a pure CENTAURO  type
  event (small $f$) is low. In fact
  there will be a shift in the $f^{\prime \prime}$ away from zero. 
  Fig.~(\ref{domain_deccay}) shows the original and modified 
  probability distributions
  $f$ and $f^{\prime \prime}$  inside a DCC domain in the model
  calculation. Here $N_{\mathrm ch}$ is only $N_{\mathrm \pi^{\pm}}$,
  the effect of detecting other charged particles along with pions
  will be discussed later. 
  Clearly one can notice the shift away from 0 to 0.1 in $f$ 
  values as a result of decay of $\pi^{0}$'s. 
  Also shown in the
  figure is the  $f$ distribution for normal events. For normal events 
  $f$ and $f^{\prime \prime}$ distributions are not very different
  because the relative population of $\pi^{0}$ 
  is the same within and outside the 
  domain and hence the loss due to decay is compensated by the gain due to
  decay.

  In order to quantify the decrease in the detectability of 
  DCC signal, we consider the following two cases.  \\

\noindent {(a) Analysis using a $\pi^{0}$ detector} \\

  In the first case we put two hypothetical 
  detectors, one for detecting $\pi^{0}$
  and other for detecting charged pions 
  with $100\%$ efficiency within one unit of $\eta$ 
  and full $\phi$ coverage. 
  The detector effects will be studied later taking realistic efficiency 
  and other parameters. The goal here is to study the $\pi^{0}$ 
  decay effect only. 
  Then we introduce DCC domains with 
  domain size $\Delta \phi$ varying from $30^{\circ}$ to $180^{\circ}$. 
  Thus set of events for a particular DCC domain size are obtained. 
  These events (without $\pi^{0}$ decay) and having DCC domain are then 
  analyzed using the DWT 
  method to obtain the FFC distribution. The strength parameter
  ($\zeta$) is calculated from Eqn.~(\ref{str_para}), with $s_{\rm X}$
  corresponding to the rms deviations of the FFC distributions of 
  DCC-type events.  \\

\noindent {(b) Analysis using an ideal photon detector} \\

  In the second case we replace the $\pi^{0}$ detector 
  with an ideal photon detector having $100\%$ photon detection efficiency
  and no contamination of charged particles. 
  The  $\pi^{0}$'s are allowed to decay. 
  Then we carry out a similar analysis as mentioned above to 
  obtain the strength values. 

  From the two strength values for each DCC domain size, 
  we calculate the percentage ($\%$) decrease in
  strength value of the second case compared to the first case.
  Fig.~(\ref{deccay}) shows the $\%$ decrease in strength of DCC signal 
  due to decay effect for various domain sizes in $\Delta \phi$.
  The results show that the decrease in 
  strength of the signal is more for smaller domains of DCC in comparison to
  the larger domains. 

  For each DCC domain size, two cases have been studied : (a)   
  all the events have DCC-type fluctuations and (b) 
  20$\%$ of the events have DCC-type fluctuations introduced.
  This is done to see the effect of number of events being DCC-type.
  We find that the results are similar 
  for both $100\%$ and $20\%$ of  events being DCC type.
\begin{figure}
\begin{center}
\includegraphics[scale=0.4]{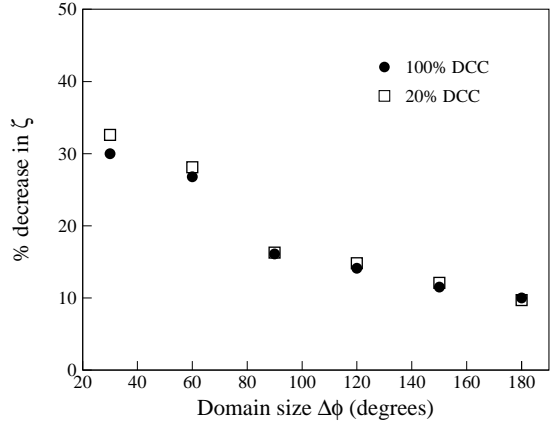}
\caption {\label{deccay}
Percentage ($\%$) decrease in the strength of DCC signal
due to the effect of decay of $\pi^{0}$ as a function of domain size 
$\Delta \phi$.
}
\end{center}
\end{figure}

\subsection{Effect of multiple domains of DCC}
\label{sec:3.1}

  In heavy-ion collisions there is a possibility that more than one domain
  of DCC will be formed with chiral condensate pointing in independent 
  direction for each of the domain. It may be mentioned that the precise
  dimension of a DCC domain is still under debate~\cite{Kaga}.
  In earlier theoretical calculations~\cite{amado} it has been shown 
  that the DCC signal approaches toward pure 
  normal case  with increase in number of domains.
  In case of multiple domains, the observed neutral pion 
  fraction $f$, which is 
  predicted to follow Eqn.~(\ref{f_prob}) for a single DCC domain, 
  will tend toward Gaussian as per the central limit theorem.
  This makes the dis-entanglement of DCC signal from that of normal events 
  difficult. Subsequent calculations~\cite{Li} suggest that the situation 
  is better for multiple domains provided one of the domains is predominant. 

  Here we attempt to quantify the effect of presence of multiple domains 
  with the help of the DCC model described earlier and using the 
  multi-resolution DWT method.
  For simplicity and in view of limitations of the model in terms of 
  introducing DCC 
  at the freeze-out stage, we have placed varying number of DCC domains 
  in an event
  depending on the domain size in $\Delta\phi$. All the domains are assumed 
  to extend 1 unit in pseudo-rapidity. For example, we can place a 
  maximum of four domains of $\Delta\phi = 90^{\circ}$, without any overlap
  among themselves. We have also carried out the analysis by placing upto
  six DCC domains in an event with  $\Delta\phi = 30^{\circ}$.
  Care is taken in placing the domains randomly so that no two domains overlap.
  The $f$ value of each domain in an event
  is randomly chosen following the probability distribution given in 
  Eqn.~\ref{f_prob}.
  The aim is to see how the signal changes in terms of the strength 
  parameter as given in  Eqn.~(\ref{str_para}).

  Fig.~(\ref{multidomain}) shows the 
  variation of  strength of DCC signal with number of domains in an event
  for two domain size of $\Delta\phi = 90^{\circ}$ and $30^{\circ}$.
  We observe that the strength of the DCC signal increases as the number 
  of domains increases and starts saturating for larger number of domains. 
  The increase in strength value with increase in number of domains 
  is because the multi-resolution event-by-event analysis
  based on DWT method has been able to pick up signals by looking 
  at bin to bin (in $\phi$) fluctuations in each event. 

  This result is not in contradiction with earlier theoretical 
  observations~\cite{amado,Li}. If the multiple DCC domains formed in the
  initial stages, in the course of their evolution, move towards the same
  part of phase space covered by a detector, then the strength of the signal
  will reduce. However if the phase space separation is maintained in evolving
  DCC domains, then the strength will increase as found in the 
  present analysis.
\begin{figure}
\begin{center}
\includegraphics[scale=0.4]{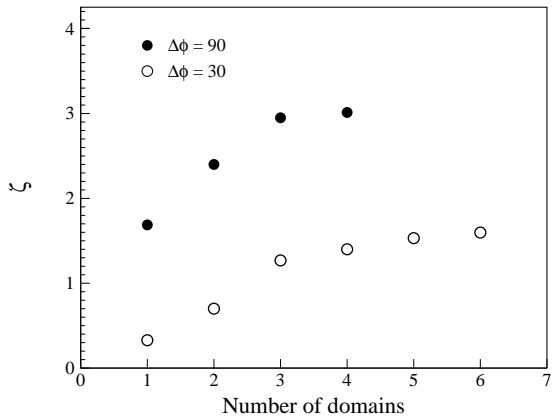}
\caption {\label{multidomain}
Variation of the strength of DCC signal
with increase in number of DCC domains in an event. 
All the events are DCC events.
}
\end{center}
\end{figure}

\section{Effects due to detector limitations}
\label{sec:4}

The detector limitations are basically those related to efficiency and purity
of particle detection and the acceptance in ($\eta$, $\phi$). 
The effect of efficiency of particle detection 
is trivial, higher the efficiency of particle detection, more reliable is 
the search for DCC. Similarly the role of higher acceptance in $\eta$ and 
full azimuthal coverage can be hardly over emphasized (lower acceptance 
effectively reduces the total multiplicity of observable DCC pions and hence
reduces the strength of the DCC signal). If the acceptance in $\eta$ is 
sizable, one can attempt DWT analysis using bins in $\eta$ for limited 
azimuthal coverage, which will be complementary to the present analysis.

The purity of the photon sample, as measured in a detector like
PMD, is around $60\% - 70\%$, whereas that of charged particle samples 
measured  using detectors like FTPC or FMD is quite high ($\ge 95\%$).
Hence we discuss here only the effect of purity of photon sample.

\subsection{Effect of neutral pion fraction on the purity of photon sample}
\label{sec:4.1}

\begin{figure}
\begin{center}
\includegraphics[scale=0.4]{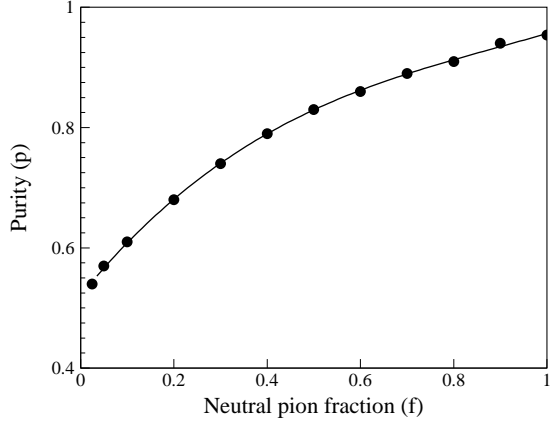}
\caption {\label{pur_f}
 Variation of purity of the photon sample 
 as a function of neutral pion fraction. For $f$ = 0.33 (generic event)
purity is $\sim$ 0.74.
}
\end{center}
\end{figure}

It is of interest to study the effect of the DCC neutral pion fraction on the
purity of photon sample. Consider a case where the neutral pion fraction 
($f$) for a DCC domain has a value of $1$. This corresponds to a case
where there are no charged pions. This will lead to greatly reduced number of 
charged particles  falling on the photon detector.
So the purity of the photon sample will be high. 
The variation of purity with neutral pion fraction can be investigated in the
following manner. 

The number of $\gamma-like$ hits on the 
photon detector can be written as
\begin{equation}
N_{\gamma-like} = \epsilon N_{\gamma} + c N_{ch}
\label{a_def}
\end{equation}
where the first term denotes the contribution of actual photons and the
second term denotes charged particle contamination.
$c$ is the fraction of $N_{ch}$ on the PMD acceptance treated as
contamination to the detected photon sample.
 
We assume that the fraction $c$ is given by a normal distribution with mean 
of $15\%$ and $\sigma$ of $5\%$, the percentage being taken with respect to 
the total charged particles within 
the acceptance of the photon detector. This is a reasonable number, 
considering that the converter thickness of the preshower detector is $\sim$
$10\%$ of an interaction length and some interactions lead to 
multiple clusters (the effect of overlapping clusters is ignored).
DCC-type fluctuations are introduced for~$fixed$~$f$~values
and domain size $\Delta \phi = 360^{\circ}$ and $\Delta \eta = 1$ 
in a set of event. Several such sets of events were generated for $f$ values 
varying from 0.025 to 1.0. The purity of photon sample is then calculated for 
each set of events with a~$fixed$~ neutral pion fraction value.

The results are shown in Fig.~(\ref{pur_f}). One clearly sees that the
purity of photon sample increases with the increase in the neutral pion
fraction value. However the purity does not reach a value of $1$ for 
$f$ = $1$,  because there are charged particle 
other than $\pi^{\pm}$ falling on the detector. Similarly it never reaches a 
value of zero for $f$ = 0, 
as there are $\gamma$'s from $\pi^0$'s decaying outside the 
detector acceptance and from other sources.

\subsection{Effect of charged particle contamination in PMD}
\label{sec:4.2}

\begin{figure}
\begin{center}
\includegraphics[scale=0.4]{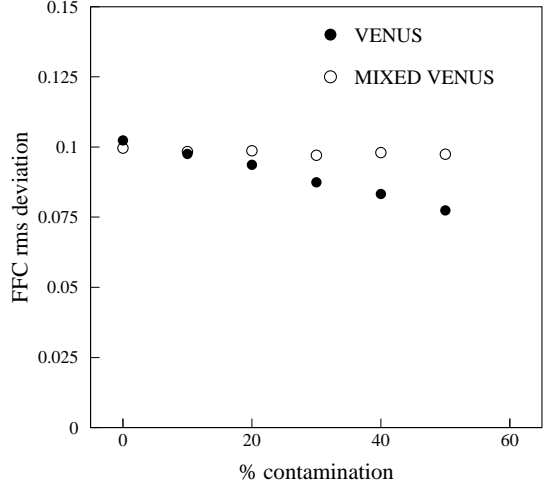}
\caption {\label{conta_var}
Variation of rms deviation of FFC distribution for pure
VENUS events with increase in $\%$ of contamination in the photon
detector. The variation of rms deviation of FFC distributions of
mixed events constructed form VENUS are also shown.
}
\end{center}
\end{figure}

  Photon measurement in heavy-ion collisions has the problem of
  charged particle contamination. 
  The charged particles detected as photons have a
  correlation with those detected in the charged particle 
  multiplicity detector. This additional correlation suppresses 
  the anti-correlation between the photons and charged particles that 
  arises due to DCC formation.
  
  The effect of such a correlation has been studied with the help of VENUS (no
  DCC) events and a set of mixed events generated from them. 
  The percentage of charged particle contamination was varied 
  from $0\%~to~50\%$. 
  The event-by-event $N_{\gamma-like}$ was kept the same for all the 
  cases. The result of the DWT analysis on these events is shown in 
  Fig.~(\ref{conta_var}). There is a decrease in the 
  rms deviation of the FFC distribution for pure VENUS events with 
  increase in the $\%$ contamination in the photon sample. 
  Thus presence of any additional anti-correlation due to DCC-like 
  effect will have to overcome the opposing correlation effect due to 
  contamination in order to be observed. 

  Mixed events form an important tool for comparison with data~\cite{WA98-12}
  in order to draw any conclusion regarding the presence of fluctuations. 
  The mixed events were 
  generated by mixing hits in the photon detector and charged particle 
  detector separately from different events and ensuring that no two 
  hits in a mixed event came from the same real event. 
  A realistic two-track resolution was assumed for the two 
  detectors~\cite{wa98nim,Lin} and this was taken into account in the
  event-mixing process.
  The mixed event 
  $N_{\gamma-{\mathrm like}}$--$N_{\mathrm ch}$
  distribution is constructed to be identical
  to the $N_{\gamma-{\mathrm like}}$--$N_{\mathrm ch}$
  distribution of the VENUS events for the full acceptance region 
  of the detectors. This is done by producing one mixed event for 
  each real event with the same multiplicity 
  of $N_{\gamma-{\mathrm like}}$ and $N_{\mathrm ch}$ pair as in the
  VENUS event.  This preserves in detail the 
  $N_{\gamma-{\mathrm like}}$--$N_{\mathrm ch}$ correlation.

\begin{figure}
\begin{center}
\includegraphics[scale=0.4]{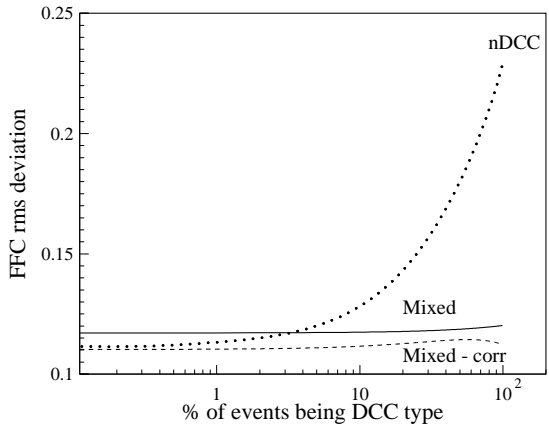}
\caption {\label{conta_var2}
 Variation of rms deviation of FFC distribution of DCC-type
  events as a function of $\%$ of events being DCC-type. Also shown
  are the results for corresponding mixed events and mixed events
  corrected for contamination effect. 
}
\end{center}
\end{figure}

  Fig.~(\ref{conta_var}) also shows the effect of contamination on
  mixed events generated from the VENUS events. 
  It is found that the rms 
  deviations of the FFC distributions for the mixed events remain independent
  of the level of contamination. The mixed events 
  appropriately break the additional 
  correlation due to charged particle contamination.

  The effect of charged particle contamination can be corrected by 
  knowing the level of contamination in photons for a given detector. 
  Fig.~(\ref{conta_var2}) shows the variation of
  rms deviation of FFC distribution for DCC-like events with increase in
  $\%$ of events being DCC-type in a given ensemble of nDCC events.
  It also shows the variation of rms deviation of 
  FFC distribution of corresponding mixed events constructed as described 
  above. The rms deviations of the FFC distribution of mixed events  are
  found to be independent of $\%$ of events being DCC-type. This is along
  the expected lines. But for lower $\%$ of events being DCC-type it 
  is above that of the parent sample of nDCC events. 
  This is because the $N_{\gamma- \mathrm
  like}-N_{ch}$ anti-correlation due  to DCC-type effect is not sufficient 
  to overcome the correlation between $N_{ch}$ and
  charged particle detected as contamination in the photon detector. 
  However as the $\%$ of events being DCC-type increases, the DCC-type 
  effect dominates. 
  The effect of the contamination can be corrected.  
  This is done by taking into account the difference in 
  rms deviation of normal events and mixed events. 
  The rms deviation of FFC distribution of corrected mixed events are 
  shown in Fig.~(\ref{conta_var2}).

The results from above two subsections indicate that it is better to look 
for anti-CENTAURO (photon excess) events in studies using 
photon and charged particle multiplicity detectors.
This is because, the effect of decay is primarily to shrink the $f$ 
distribution from the lower $f$ side and high value of $f$ leads to a
reduced effect of charged particle contamination on DCC search.

\subsection{Effect of $p_{T}$ information of charged particles and photons}
\label{sec:4.3}

  It is expected that  $p_{T}$ information of particles would be very
  helpful in DCC search. This would also enable one to  verify the various 
  predicted features of DCC formation, such as, DCC-pions are 
  low $p_{T}$ pions and DCC formation may lead to low $p_{T}$ enhancement 
  in pions~\cite{Kaga}. 

  In order to show the utility of $p_{T}$ information we carried out
  the following study. Here we assume that all charged particles with
  $p_{T}$ greater than $50$ MeV/c are detected ( the $p_T$ acceptance of
  preshower detector like PMD extends down to $30$ MeV/c~\cite{wa98nim}). 
  The $p_{T}$ 
  resolution for the charged particle detector was taken as 
  $\frac{\Delta p_{T}}{p_{T}}$ = $0.2$~\cite{ftpc}. These realistic parameters
  are taken from a typical experiment for DCC search, as in STAR.
  Since DCC pions are believed to have 
  low $p_{T}$, in our simulation we introduce DCC-type
  fluctuations in pions with $p_{T}$ $\le$ 150 MeV/c. 
  Then a DWT analysis is carried out.
  The sample function in the DWT analysis is modified, such that  
  $N_{\mathrm ch}$ taken corresponds to those having $p_{T}$ $\le$
  $150$ MeV/c.

\begin{figure}
\begin{center}
\includegraphics[scale=0.4]{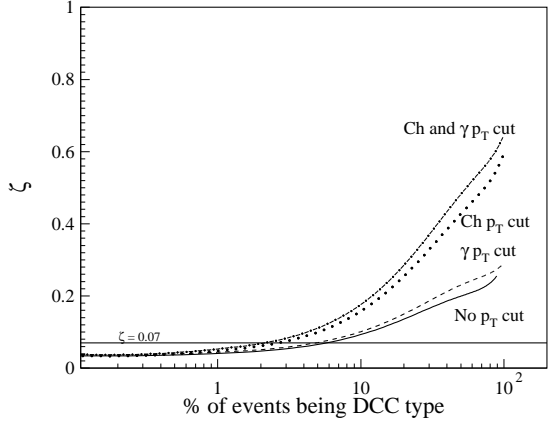}
\caption {\label{lowpt}
  Variation of $\zeta$ as a function of $\%$ of events being
  DCC-type, for charged particle detector with $p_{T}$ information 
  and  photon detector without $p_{T}$ information (dotted line), 
  for photon detector with $p_{T}$ information and
  charged particle detector without $p_{T}$ information (dashed line)
  and for both charged particle and photon detector with $p_{T}$ 
  information (dot-dashed line).
  Also shown are the corresponding results with both charged particle
  detector and photon detector without $p_{T}$ information (solid line).
  The horizontal straight line indicates the statistical error on $\zeta$.
}
\end{center}
\end{figure}
  In Fig.~(\ref{lowpt}) we show the strength, $\zeta$, of the DCC signal as a 
  function of percentage ($\%$) of events being DCC-type. For comparison 
  also shown is the corresponding $\zeta$ where the charged particle 
  detector has no $p_{T}$ information. We note that the absolute 
  value of $\zeta$ is lower compared to those presented in previous sections.
  This is because here we have taken only low $p_{T}$ pions as DCC pions in 
  a given domain, whereas earlier all the pions in the domain were 
  considered to be
  DCC-type. This reduces the strength $\zeta$ as a result of reduction in 
  multiplicity. The statistical error on $\zeta$ for the present case is 0.07.

\begin{figure}
\begin{center}
\includegraphics[scale=0.36]{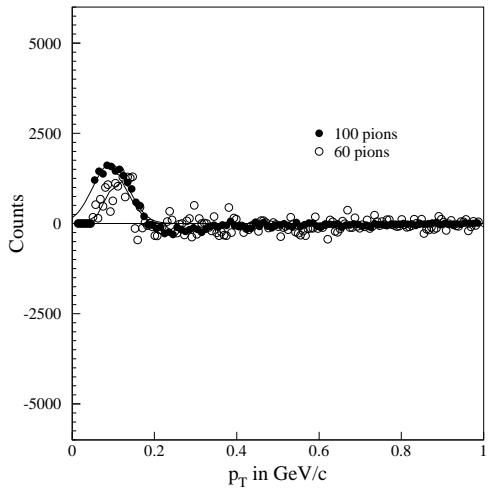}
\caption {\label{pt_enhance}
 Resultant $p_{T}$ distribution of charged particles with 
$20\%$ of the events being DCC-type, obtained by subtracting the
$p_{T}$ distribution of charged particles within $2~\sigma$ of the
FFC distribution from those beyond. Two cases are presented, additional
low $p_{T}$ DCC pions being 60 and 100.
The statistical errors are within the symbol size.
}
\end{center}
\end{figure}

  Clearly one can see the increase 
  in strength of the signal with the use of $p_{T}$ information. We carry out
  the same analysis assuming that all photons have $p_T$ information
  while the charged particles do not have any $p_{T}$ information.
  The results are also shown in Fig.~(\ref{lowpt}). One finds that
  although there is an increase in signal strength, it is more for event sample
  having a larger fraction of DCC events and much less compared to having
  $p_T$ information for charged particles. For the case where the analysis
  is carried out assuming that both photons and charged particles have 
  $p_{T}$ information, one finds that the results are close to those obtained 
  for charged particle detector with $p_{T}$ information, except for events
  having higher percentage of DCC type events.

  It is also believed that DCC formation may lead to low $p_{T}$ enhancement.
  This hypothesis can be checked if the $p_{T}$ of charged particles is 
  measured. In order to incorporate this effect in simulation, a number of 
  low $p_{T}$ pions ($p_{T}$ $\le$ 150 MeV/c), 
  generated according to DCC probability given in Eqn.~(\ref{f_prob})
  and having uniform $p_{T}$ distribution were added within 
  the chosen domain  on top of the existing pions of a normal 
  event~\cite{dccstr}. 
  The number of pions to be added depends on the size of DCC domain and
  energy density of the domain. Assuming that the energy density within a
  DCC domain is about $50$ MeV/$fm^{3}$ and domain radius of the order
  of $3-4$ fm ~\cite{Kaga} we have performed simulations taking $100$ and $60$
  additional low $p_{T}$ pions. As a test case
  we have taken a sample with $20\%$ of the events 
  having a DCC domain of size $\Delta \phi$ =$90^{\circ}$ and 
  $\Delta \eta$ = 1. 

  The event sample is analyzed by the DWT method to 
  obtain the FFC distribution, which was found to be near Gaussian for 
  the present case. 
  We divide the FFC distribution into two parts, (a) one within $\pm 2 \sigma$
  of the mean and (b) another beyond this. We obtain the $p_{T}$ 
  distributions of the two sets of events.
  Then we subtract the histogram corresponding to events within 
  $2\sigma$ of the  FFC distribution from those beyond, 
  after proper normalization.  The resultant  spectrum is 
  shown in Fig.~(\ref{pt_enhance}).  

  Clearly one sees that if DCC results in low $p_{T}$ enhancement then 
  following the above analysis technique in data, one should be able to observe
  such an effect. The effect is more for case of 100 additional DCC pions 
  compared to case of  60 additional DCC pions.

\section{Signal and Background of DCC events}
\label{sec:5}

\begin{figure}
\begin{center}
\includegraphics[scale=0.4]{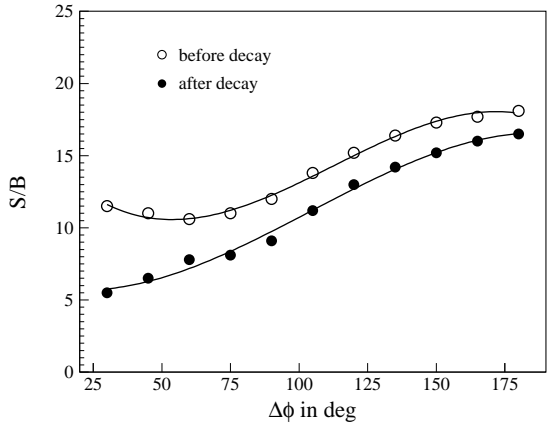}
\caption {\label{deccay_sb}
Signal (S) to Background (B) ratio of DCC signal
before and after the decay of $\pi^{0}$, as
a function of DCC domain size.
}
\end{center}
\end{figure}

The signal to background ratio of DCC events defined in section 2.4
gets modified due to $\pi^{0}$ decay (which leads to shrinking of the
$f$ distribution for DCC events) and also due to the effect of charged
particle contaminants in the photon detector. In this section we
first discuss the effect of $\pi^{0}$ decay and then we
discuss the effect of various detector effects on the signal to
background ratio of DCC events.

An analysis of the events as described in section 2.1, 
to evaluate the signal to background ratio gives similar results 
as obtained from the DWT method. Fig.~(\ref{deccay_sb}) shows the 
$S/B$ values before and after decay as a function of size of the DCC
domain in azimuth. One clearly observes that the decrease in 
$S/B$ value for smaller domains is more  compared to that for larger domains.

\begin{figure}
\begin{center}
\includegraphics[scale=0.4]{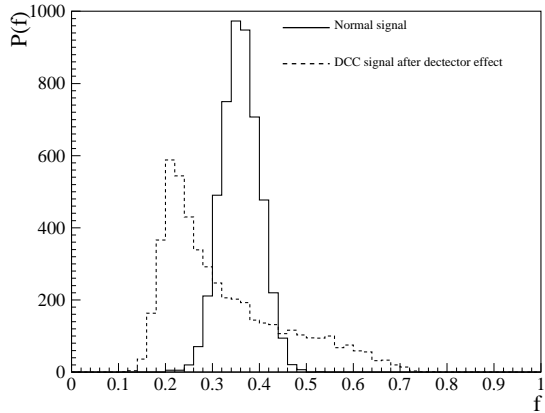}
\caption {\label{domain_detector}
Typical $f$ distributions inside a DCC domain after detector
effects, such as charged particle contamination and realistic efficiency
of photon and charged particle detectors. Also shown for comparison is the 
$f$ distribution of the corresponding normal events.
}
\end{center}
\end{figure}

The efficiency along with the charged particle contamination in photon
sample as discussed in previous sections, 
causes a further shrinkage in the $f$ values 
inside a typical DCC domain as shown in Fig.~(\ref{domain_detector}). 
This is in addition to those due to $\pi^{0}$ decay shown 
in Fig.~(\ref{domain_deccay}).  The $f$ distribution for normal 
events is now wider.

Signal to background ratio for typical detector parameters are shown in  
Figs.~(\ref{sb_detector_domain}) and (\ref{sb_detector_event})
as a function of DCC domain size and percentage of events being
DCC-type for a given domain size respectively. From 
Fig. ~(\ref{sb_detector_domain}) one observes that the $S/B$
ratio increases with increase in domain size. If one separates
the S/B ratio for photon excess and charge excess events, then one observes
that S/B ratio for photon excess events is larger for smaller DCC domains.

\begin{figure}
\begin{center}
\includegraphics[scale=0.4]{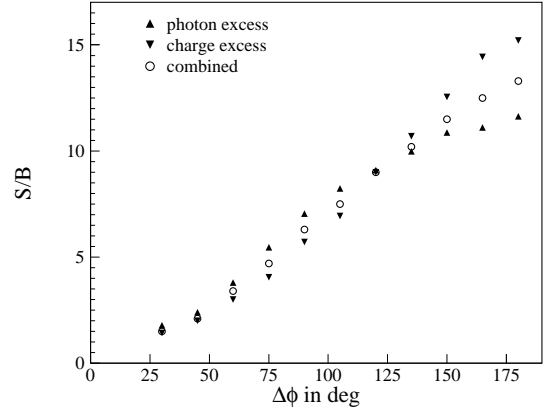}
\caption {\label{sb_detector_domain}
The variation of signal(S) to background (B)
ratio as a function of DCC domain size. 
In the event sample all events are DCC type and the results are obtained 
after including decay and detector effects. 
The S/B values for photon excess (higher $f$ values) and charge excess (lower $f$ values) cases are also shown.
}
\end{center}
\end{figure}
Fig.~(\ref{sb_detector_event}) shows the $S/B$ ratio 
as a function of percentage of events being DCC-type for a 
DCC domain of $90^{\circ}$ in azimuth. One observes that the S/B
ratio increases with increase in the number of events being DCC-type.
One also notices that the S/B ratio of photon excess events 
is much larger than those for charge excess events.

Although these results are along expected lines (the signal increases
as the number of DCC pions increases and also as number of events 
with DCC-type fluctuations increases), it clearly shows that it is more 
advantageous to
look for photon excess events.
\begin{figure}
\begin{center}
\includegraphics[scale=0.4]{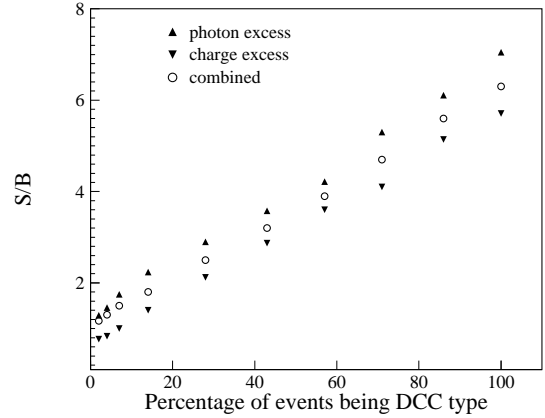}
\caption {\label{sb_detector_event}
The  variation of signal (S) to background (B)
ratio as a function of percentage of events being DCC type in a given
sample of events. 
Here the DCC domain chosen is $90^{\circ}$ in $\phi$ and the results are 
obtained after taking the detector effects into account.
The S/B vales for photon excess (higher $f$ values) and charge excess (lower $f$ values) cases are also shown.
}
\end{center}
\end{figure}

\section{Discussion and Summary}
\label{sec:6}

  The results presented here describe the limitations and the 
  possibilities of DCC search in heavy-ion collisions. We have 
  discussed various factors that affect the DCC signal 
  (fluctuation in $f$) using a simple DCC model, a multi-resolution 
  analysis technique based on wavelet transformation and an estimation
  of signal to background ratio.  

  Among the possibilities, contrary to the theoretical 
  results so far on multiple domains of DCC, we find that these can
  still be looked for without much loss in signal strength if the domains do
  not overlap.
  This is possible through a multi-resolution analysis carried out 
  by binning the $\eta-\phi$ phase space, as done in the discrete
  wavelet transformation method. Increase in multiplicity of the particles
  produced as one goes from SPS to RHIC and LHC energies results in
  lower statistical fluctuations in the observables analyzed.
  We have also shown that the possibility of DCC search is increased if 
  charged particle detector has $p_{T}$ information. However one finds
  that, primarily due to the decay kinematics of $\pi^{0}$ to photons, the
  effect of photon detector with $p_{T}$ information on DCC signal strength
  is not appreciable. 
  A charged particle detector having $p_{T}$ information
  along with photon multiplicity detector will be
  a better setup to look for fluctuations in primary signal of DCC,
  i.e the photon fraction, specially for the anti-CENTAURO type events. 

  One of the major limitation to the detection of DCC signal is  
  the decay of $\pi^{0}$ to photons. It also reduces the possibility for
  observing low $f$ (CENTAURO) events. 
  The other major limitation is the correlation arising due 
  charged particle contamination in the 
  photon detector. It reduces the strength of the DCC signal. However, 
  this limitation can be overcome by knowing the amount of charged 
  particle contamination in the photon detector. 
  The purity of photon sample increases for a DCC event with higher 
  $f$ value. Thus looking for higher $f$ (anti-CENTAURO) events is
  advisable. 

  The other limitations are due to efficiency,
  granularity and the finite acceptance of the detectors. Reduced acceptance
  greatly reduces the strength of the DCC signal. Thus DCC studies will
  be affected using the TPC + PHOS combination in ALICE, even though this 
  combination will be able to provide $p_{T}$ information for both photon 
  and charged particles.

  One observes that the signal to background ratio for photon 
  excess (anti-CENTAURO) events is higher than those for the 
  charge excess (CENTAURO) events for DCC domains upto 
  $120^{\circ}$ in azimuth. And for these domains the S/B ratio for 
  the photon excess events are higher compared to charge excess events 
  as a function of percentage of events being DCC-type.
  Thus it is advisable to look for photon excess events in an experimental
  event sample for looking at DCC-type fluctuations, using a photon 
  multiplicity detector and a charged particle multiplicity detector having
  particle wise $p_{T}$ information. \\

\noindent {\bf Acknowledgments : }{ One of us (B.M.) is grateful 
to the Board of Research
on Nuclear Science of the Department of Atomic Energy,
Government of India for financial support in the form of Dr. K.S. Krishnan
Fellowship.}

\end{document}